\documentclass[journal=amlccd,manuscript=letter]{achemso}
\usepackage{graphicx}
\usepackage{amsmath}
\usepackage{amssymb}
\usepackage[version=3]{mhchem} % Formula subscripts using \ce{}
\usepackage{comment}
\usepackage{microtype}
\usepackage[numbers]{natbib}
\usepackage{natmove}
\usepackage{hyperref}
\usepackage{subcaption}
\usepackage[nolist,nohyperlinks]{acronym}
\begin{acronym}[RNEMDS]
  \acro{AFM}{atomic force microscopy}
  \acro{DPN}{dip-pen nano-lithography}
  \acro{hCG}{highly coarse-grained}
  \acro{HPC}{high performance computing}
  \acro{LJ}{Lennard-Jones}
  \acro{MC}{Monte-Carlo}
  \acro{MCS}{Monte-Carlo steps}
  \acro{MD}{molecular dynamics}
  \acro{MFA}{mean-field approximation}
  \acro{MSD}{mean squared displacement}
  \acro{PMMA}{poly(methyl methacrylate)}
  \acro{PS}{poly(1-phenylethene)}
  \acro{SCFT}{self-consistent field-theory}
  \acro{SCMF}{single-chain-in-mean-field}
  \acro{SLSP}{slip-spring}
  \acro{SLLI}{slip-link}
  \acro{DPD}{dissipative particle dynamics}
  \acro{MDPD}{many-body dissipative particle dynamics}
  \acro{PI}{poly(isoprene)}
  \acro{FENE}{finite extensible nonlinear elastic}
  \acro{RCC}{Research Computing Center}
\end{acronym}

\newlength{\figurewidth}
\newlength{\multifigurewidth}
\graphicspath{figures}

\author{Ludwig Schneider}
\altaffiliation{Pritzker School of Molecular Engineering, University of Chicago, 5740 S. Ellis Ave, Chicago, IL, USA}
\author{Juan de Pablo}
\altaffiliation{Prizker School of Molecular Engineering, University of Chicago, 5740 S. Ellis Ave, Chicago, IL, USA}
\email{depablo@uchicago.edu}

\title{Entanglements via Slip-Springs with Soft, Coarse-Grained Models for Systems Having Explicit Liquid-Vapor Interfaces}

\abbreviations{CG, SLSP}

\begin{document}

\begin{abstract}
Recent advances in nano-rheology require that new methods and models be developed to describe the equilibrium and non-equilibrium properties of entangled polymeric materials and their interfaces at a molecular level of detail.
In this work we present a Slip-Spring (SLSP) model capable of describing the dynamics of entangled polymers at interfaces, including explicit liquid-vapor and liquid-solid interfaces.
The highly coarse-grained approach adopted with this model enables simulation of entire nano-rheological characterization systems within a particle-level base description.
Many-body dissipative particle dynamics (MDPD) non-bonded interactions allow for explicit liquid-vapor interfaces, and  compensating potential within the SLSP model ensures unbiased descriptions of the shape of the liquid-vapor interface.
The usefulness of the model has been illustrated by studying the deposition of polymer droplets onto a substrate, where it s shown that the wetting dynamics is strongly dependent on the degree of entanglement of the polymer.
More generally, the model proposed here provides a foundation for the development of digital twins of experimentally relevant systems, including a new generation of nano-rheometers based on nano- or micro-droplet deformation.

\end{abstract}

\section{Introduction}
Entanglements give polymeric materials their unique viscoelastic properties. Individual macromolecules can reach contour lengths of many microns, and the resulting topological constraints that arise in condensed polymeric phases lead to long-term relaxation processes that are very different from those observed in simple small-molecule liquids \cite{de1982dynamics}.
These properties are well known for bulk materials; however, the rise of nano-rheology \cite{gillies2004interaction,waigh2016advances} presents intriguing opportunities to examine the role of entanglements in settings where confinement restricts the material to length scales comparable to the size of individual molecules. Understanding the underlying rheology could in fact help nano-rheology measurements become a standard tool for the characterization of bulk rheology from ultra-small samples.
As discussed in this work, coarse-grained simulations provide a unique means to model entire nano-rheology systems, such as droplets, while retaining molecular information.
Slip-Spring (SLSP) models\cite{chappa2012translationally,uneyama2012multi,Hernandez2013Jan,Hernandez2013Aug,ramirez2017multi,sgouros2017slip,vogiatzis2017equation,masubuchi2018multichain,langeloth2013recovering,masubuchi2016multichain,megariotis2018slip,behbahani2021dynamics,schneider2021simulation,li2021dynamics,megariotis2016mesoscopic,megariotis2018slip,sgouros2017slip,hollborn2022effect} have gradually been developed to describe the dynamics of high molecular weight entangled polymeric liquids. A particle-level description ensures that the dynamics of the system are captured correctly, and slip springs are used to mimic the effects of entanglements within the context of a highly coarse-grained representation\cite{muller2011studying,tschop1998simulation,harmandaris2006hierarchical,Praprotnik08,fritz2009coarse,padding2011systematic,webb2018graph,behbahani2021dynamics,li2021dynamics,dhamankar2021chemically}. More specifically, a high degree of coarse-graining is generally accompanied by the use of soft non-bonded interaction potentials that are unable to prevent chain crossing; the artificial springs encoded in the \acp{SLSP} representation serve to reintroduce topological constraints.
Much of the literature on \ac{SLSP} models has focused on establishing their correct asymptotic behavior (\textit{e.g.} showing that the models are capable of capturing the power-law dynamics predicted by tube models \cite{doi1988the,chappa2012translationally,uneyama2012multi,ramirez2017multi}).
More recent efforts have sought to introduce systematic procedures to parameterize models for specific chemical systems or more complex molecular architectures\cite{behbahani2021dynamics,ramirez2018detailed,liang2022bottom,li2021dynamics}.
Importantly, the existing body of work on \ac{SLSP} models has been limited to the bulk properties of polymers.

In this work, we present a \ac{SLSP} model that is capable of describing entangled polymers at interfaces, including hard surfaces and explicit vapor-liquid interfaces.
To the best of our knowledge, combining \acp{SLSP} and an explicit liquid-vapor interface represents a new development that opens new avenues for comprehensive investigations of polymer nanorheology \textit{in silico}.
The explicit liquid-vapor interface eliminates the need for explict dummy particles that represent the gas phase as a structureless liquid, as proposed in previous work\cite{dreyer2022simulation}, or the need for explicit confinement\cite{michman2019controlled}, thereby lowering computational demands while improving accuracy and fidelity.

\section{Results and Discussion}

A number of existing \ac{SLSP} models used to describe shear flows rely on a \ac{DPD} thermostat \cite{groot1997dissipative,espanol1995statistical} to control temperature.
The \ac{DPD} thermostat conserves the momentum of the particles locally, thereby enabling simulation of flowing systems.

In \ac{DPD} formalism, nonbonded interactions are reduced to a simple quadratic repulsion expression
\begin{align}
    V_C(r_{ij}) = \frac{A_{ij}}{2} (1-r_{ij}/r_c)^2,\label{eq:dpd-normal}
\end{align}
where the positive parameter $A_{ij}$ controls the compressibility and phase separation of different species, with $A_{ij} > A_{ii}$.
This purely repulsive interaction neglects the attractive long-range interactions that characterize molecular systems.
For bulk systems, the pressure can be accounted for with a tail correction. However, the use of a long-range potential incurs into significant computational costs.

Instead of relying on standard \ac{DPD} thermostat, here we resort to the \ac{MDPD} potentials that were originally introduced by \citeauthor{warren2003vapor}\cite{warren2003vapor,warren2013no,zhao2021review}.
Non-bonded interactions are divided into two parts: \autoref{eq:dpd-normal} acts as an attractive potential when a negative $A_{ij}$ parameter is employed. A second many-body potential contribution is then added, which depends on the instantaneous density, $\hat \rho_i$, around the particles $i$
\begin{align}
    V_M(r_{ij}) = \frac{B}{2} (\hat\rho_i + \hat \rho_j) (1-r/r_M)^2\\
    \hat \rho_i = \sum_j \frac{15}{2\pi r_M^3}(1-r/r_M).
\end{align}
The parameter $B$ controls the strength of the repulsion, and $\rho_i$ describes the total local density around particle $i$. Note that the cutoff distance is different for the two parts of the potential $r_M \neq r_c$.

Here we use parameters $A = 40$ and $B=10$, with cut-off distances $r_c = \sigma > r_M = 0.75\sigma$. These values correspond to a state point along the liquid-vapor coexistence curve, with a density of $\rho = 7.1\pm 0.2 \sigma^{-3}$ in the liquid branch.

\begin{figure}
    \centering
    \includegraphics[width=\figurewidth]{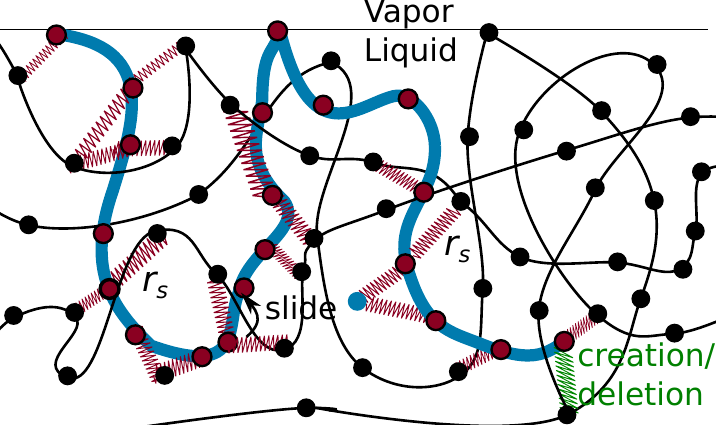}
    \caption{Schematic representation of the \ac{SLSP} model in the vicinity of a liquid-vapor interface, showing the \ac{MDPD} non-bonded interactions. Topological constraints are described by a slip spring \ac{SLSP} connecting a chain (blue) to other neighboring chains (black). The vapor phase reflects the chain contour and does not pose any additional topological constraints.}
    \label{fig:slsp-schematic}
\end{figure}

The remaining parts of the model are adopted from Ref.~\cite{chappa2012translationally, behbahani2021dynamics}.
\autoref{fig:slsp-schematic} provides a schematic representation of the \ac{SLSP} model in the presence of a liquid-vapor interface.
Polymer chains are represented with $N=64$ beads connected by harmonic bonds $V_\text{b}(b) = \frac{1}{2}k b^2$, with a spring constant of $k=16/3 k_\text{B}T/\sigma^2$.
The average bond extension is $b_0 = \sqrt{\langle b^2\rangle} = 3/4\sigma$.
Entanglements are introduced through FENE slip springs in a grand canonical ensemble: $V_\text{ss}(r) = -\frac{k_\text{ss}r_\text{ss}^2}{2} \log(1-(r/r_\text{ss})^2)$ with $k_\text{ss} = k$ and $r_\text{ss} = \sigma$.
The average number of \acp{SLSP} is controlled by a chemical potential $\mu$.
We describe the system via a fugacity $z = e^{\mu/k_B T}$ that is proportional to the average number of \acp{SLSP}: $\langle n_{ss} \rangle \propto z$\cite{chappa2012translationally}.
We integrate the system in time with a time step $\Delta t = 10^{-3}\tau$, implemented in HOOMD-blue version 2.9.7 with custom plugins\cite{anderson2008general,phillips2011pseudo,glaser2015strong,anderson2020hoomd}.
We update the \ac{SLSP} configuration with a Slide move every $10^2$ time steps to describe reptation, and with creation and deletion moves at the chain ends every $10^3$ time steps to describe tube renewal.
The additional effective attractions induced by \acp{SLSP} are compensated by an additional potential of the form ~\cite{chappa2012translationally}
\begin{align}
  V_{\text{comp}}(\boldsymbol{r}) = k_\text{B}Tz\exp(-\beta V_\text{ss}(\boldsymbol{r})). \label{eq:Vc}
\end{align}

This compensating potential is particularly important in the case of a liquid-vapor interface.
In bulk systems, past studies have sought to counter the attractive interaction of the \acp{SLSP} with a pressure correction\cite{ramirez2017multi, langeloth2013recovering, megariotis2018slip}. However, for liquid-vapor systems, the pressure of the liquid phase must be corrected during the simulation run without altering the liquid-vapor coexistence.
We therefore include an explicit compensating potential, as described in Ref.~\cite{chappa2012translationally}.
The importance of the compensating potential is highlighted in the Supporting Information Figure S1: without it, the liquid-vapor interface position and shape change depending on the number of \acp{SLSP}.
The springs are designed to only alter the dynamics of the system, and not the thermodynamics.
The static and dynamic properties can be adjusted independently, which is important for the top-down parameterization of the model required for description of experimentally relevant systems.
In the presence of an explicit liquid-vapor interface, a missing compensating potential cannot be replaced by a pressure correction.

We also introduce a substrate into our system, which we model with a \ac{LJ} type wall and parameters $\epsilon_\text{LJ} = 10 k_\text{B} T$, $\sigma_\text{LJ} = 3/4\sigma$. The cutoff distance is given by $r_\text{LJ} = \sigma$.
Using the \ac{MDPD} model, as opposed to the more common \ac{DPD} model, automatically creates a realistic packing close to the interface and compensates for the "missing neighbor" effect. As noted in~\cite{kumar2022wall}, it is not explicitly necessary. The Supporting Information Figure S1 illustrates this packing: the high degree of coarse-graining requires a flat profile. Note the absence of any packing peaks close to the interface, since atomistic packing only appears at smaller length scales.

\begin{figure}
    \centering
    \includegraphics[width=\multifigurewidth]{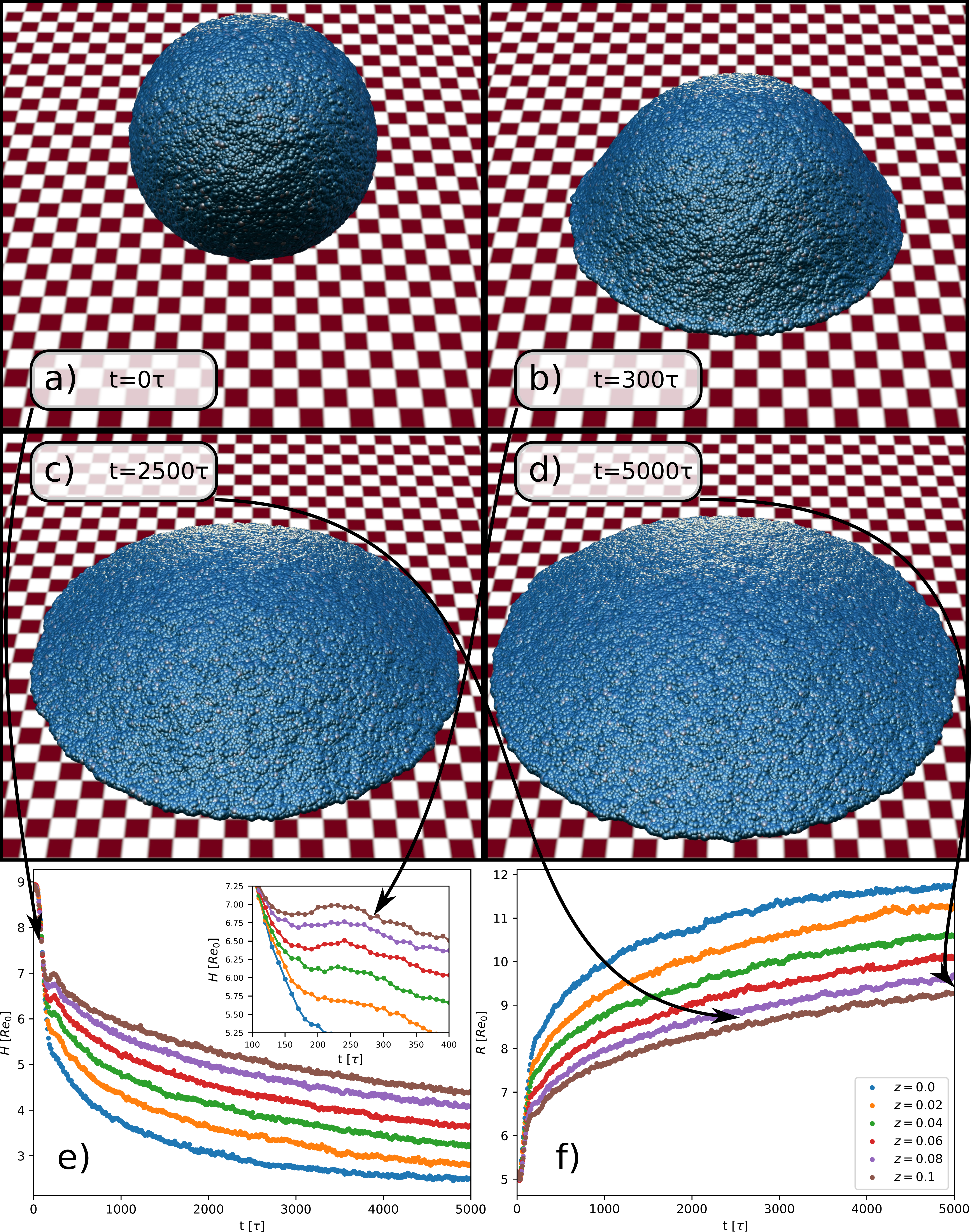}
    \caption{Deposition of an entangled ($z=0.1$: $n_\text{SS}/n\approx12$) droplet on an attractive surface. We model the deposition by quenching the surface interaction $\epsilon_\text{LJ} (1\rightarrow 8.5) k_B T$. a-d) Time evolution of the droplet. The height e) and radius f) of the droplets are shown for different degrees of entanglement.}
    \label{fig:droplet-radius}
\end{figure}

As a representative application, in what follows we use the model to quantify the deposition of a polymer droplet onto a substrate.
Such a procedure can be performed experimentally by \ac{DPN} \cite{liu2020evolution} nanolithography.
Experiments rely on entangled, high molecular weight polymers ($M_w \approx 5\cdot10^5 $g/mol), and we are not aware of other, existing models capable of describing the droplet deposition process for highly entangled polymeric liquids or inks.
We mimic such a process by preparing a droplet of realistic size ($n=12867$ chains) on a low energy \ac{LJ} $\epsilon_\text{LJ} = k_B T$ substrate.
For a low surface energy, the surface tension dominates the structure of the equilibrium droplet, which adopts an almost spherical shape, thereby mimicking the initial droplet deposition (\autoref{fig:droplet-radius}a).
From there, we quench the surface energy to $\epsilon_\text{LJ} = 8.5 k_B T$, thereby wetting the surface at equilibrium.
We examine the deformation process that ensues as a function of the entanglement density $z \propto n_{ss}$.

\autoref{fig:droplet-radius}e-f) provides a pictorial representation of this deformation process as a function of time. We show the radius of the droplet and its height on the substrate.
We observe that, initially, the droplet deforms rapidly, in a manner that is independent of the number of topological constraints \acp{SLSP}.
In this initial process, the lower half of the droplet deforms to maximize its contact with the surface (\autoref{fig:droplet-radius}b) without deforming its core.
As expected, a more entangled system (i.e., for larger $z$), has a less compliant core.
Strongly entangled droplets exhibit a "rebound" effect (inset of \autoref{fig:droplet-radius}e).
The entangled core behaves like a rubbery network that is stretched by attraction to the surface.
An elastic rebound occurs, and the height and radius time evolution slopes are temporarily reversed.
After this rebound, sufficient time elapses for the entanglements to relax, and the droplet's core is able to deform in a liquid-like manner.

The deformation of the core relaxes considerably more slowly because chain conformations must be reordered.
In the entangled case, this requires the tube-renewal to adjust to the new shape of the droplet.
However, since \acp{SLSP} only affects dynamics, we expect the equilibrium shape of the droplet to be independent of the entanglement density.% unless entanglements get trapped by the droplet shape.

\begin{figure}
    \centering
    \includegraphics[width=\multifigurewidth]{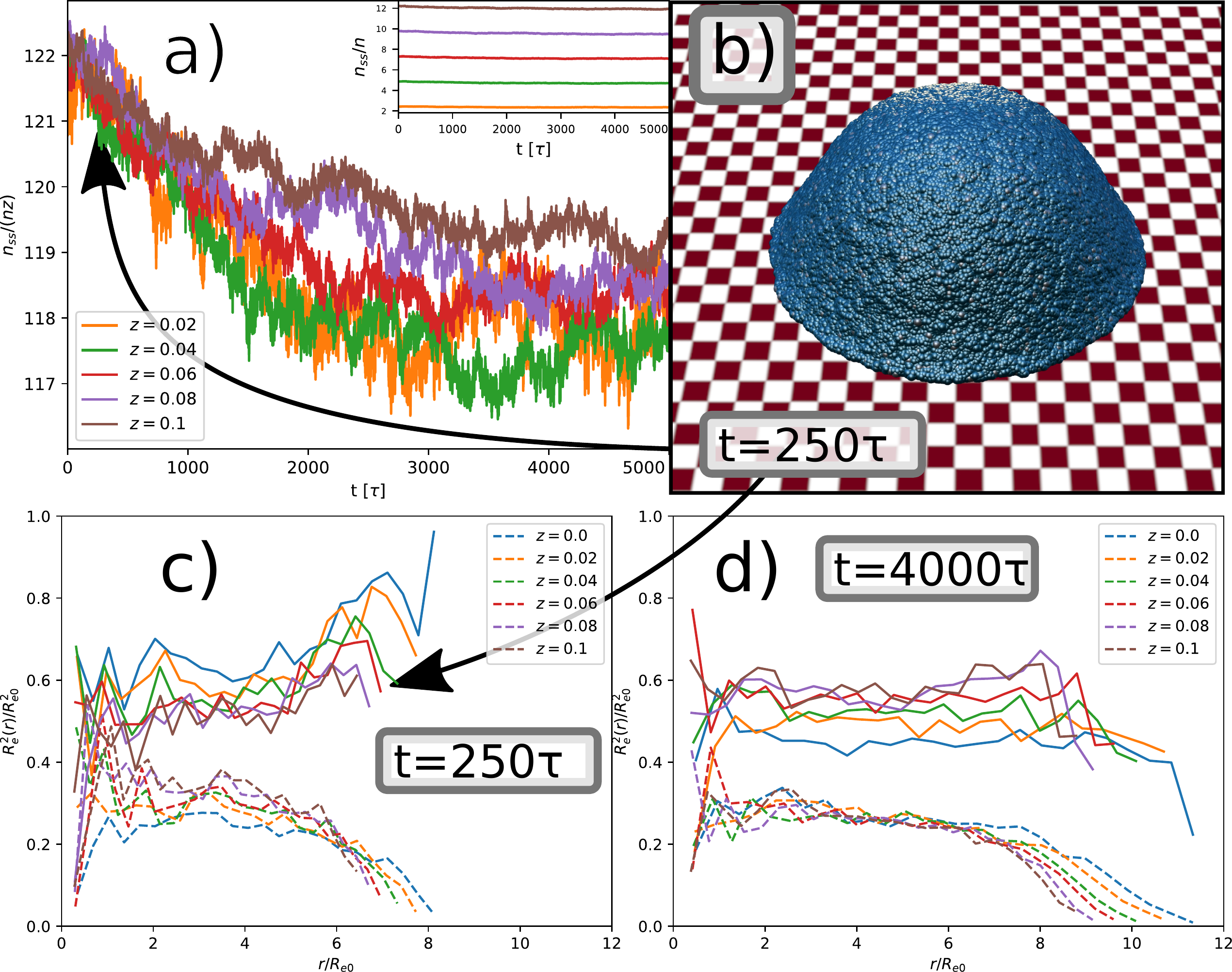}
    \caption{(a) Number of \acp{SLSP} as the droplet deforms with time. The number of \acp{SLSP} is normalized by, $z$, and the curves collapse initially as expected. Later, the number of entanglements decreases faster for lighter entangled systems because the deformation of the droplet (b) is faster too (\autoref{fig:droplet-radius}). The inset shows the unnormalized number of \acp{SLSP} to show the small absolute change and the absolute number of \acp{SLSP} per chain as function of the fugacity $z$. (c) Shows the chain conformations demonstrated by parallel (continuous) and perpendicular (dashed) component of the end-to-end vector squared, $R_e^2$, in relation to the substrate surface as a function of the droplet radius $r$.}
    \label{fig:nss}
\end{figure}

As the shape of the droplet evolves, the surface area-to-volume ratio increases.
Any chain close to an interface can only become entangled with other chains on the liquid side of the interface (\autoref{fig:slsp-schematic}). Therefore, we expect fewer entanglements as the surface area increases.
In \autoref{fig:nss} we observe that, as expected, the number of \acp{SLSP} decreases over time.
Consistent with the slower deformation of highly entangled droplets, the relative decrease in the number of \acp{SLSP} is also slower.
As the equilibrium droplet shape is approached, the deformation slows and the number of \acp{SLSP} also approaches a plateau.
In general, this effect is not strong; the change in the number of entanglements is systematic, but of order $1\%$.

The droplet deposition process presented here for a particle-based model provides insights into the molecular conformations during the deposition.
Important information can be extracted by examining the components of the end-to-end vector, $\boldsymbol{R}_e$, of the polymer molecules.
With the substrate breaking the symmetry, a decomposition into the components parallel to the surface and perpendicular to the surface becomes of interest.
The \acp{SLSP} model reveals important differences in the dynamics with and without the effects of entanglements.
\autoref{fig:nss}c) shows both components of $\boldsymbol{R}_e$ inside the droplet as a function of radius $r$ at an early stage $t=250\tau$, when the droplet has just made contact with the surface, but the shape has not yet fully relaxed.
We observe that the parallel component of $\boldsymbol{R}_e$ is larger than the perpendicular component.
This is expected, since the droplet is already in good contact with the substrate and the chain conformations are reflected from the substrate\cite{silberberg1982distribution}.
When compared to the center of the droplet, near the edge  we observe that the parallel component increases while the perpendicular component decreases.
This is explained by the spread of the droplet; in the snapshot of  \autoref{fig:nss}b), we see a lip that forms around the core of the droplet that spreads out first.
This lip is responsible for the change in molecular conformation: the chains stretch out, thereby maximizing the surface area.
Interestingly, this effect is strongest in the unentangled system ($z=0$).
The reason is that with fewer topological restrictions chain dynamics are faster.
At later times, (\autoref{fig:nss}d), the effect is reversed,
and the parallel component of the end-to-end vector is smaller at the edge of the droplet.
In this stage, the droplet is fully relaxed, and chain conformations are closer to equilibrium with one another.
Here, the lip has also relaxed, making the droplet and chain conformations uniform throughout the entire sample.
Just at the edge of the droplet, we observe that the chain extensions are getting smaller towards the edge, but this is consistent for both the parallel and perpendicular components, indicating the overall thin height of the droplet.
We can also see that the strongly entangled system does not relax the parallel chain conformation, because the topological constraints require a full tube renewal for conformational relaxation.
This also explains why shape deformation is protracted for highly entangled droplets.

\begin{figure}
    \centering
    \includegraphics[width=\multifigurewidth]{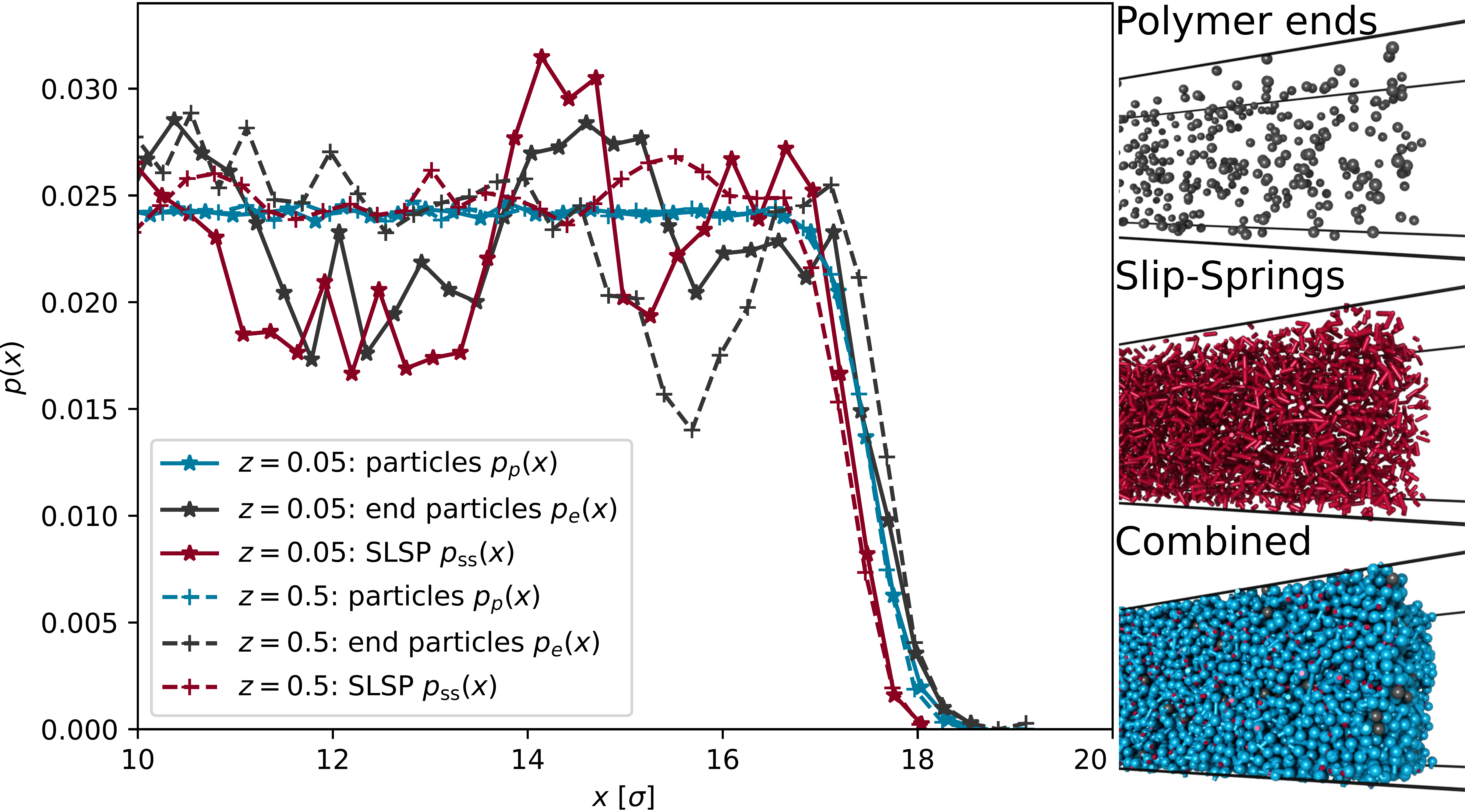}
    \caption{Normalized probability distributions of particle, polymer-end-particles and \acp{SLSP}. A system snapshot is presented for reference. Continuous lines describe the moderately entangled system with $z=0.05$, while dashed-lines correspond to a highly entangled system $z=0.5$. The right hand-side visualizes the same density profiles with simulation snap-shots that show the spatial position of polymer ends, \ac{SLSP} and all particles.}
    \label{fig:flat-profile}
\end{figure}

Our findings raise the question of whether \acp{SLSP} interact with the liquid-vapor interface.
To address this issue, we examine the probability of finding a \acp{SLSP} close to the interface.
\autoref{fig:flat-profile} shows the probability distribution relative to the density distribution, along with the density distribution of polymer-end-particles in a system having a planar liquid-vapor interface.
We observe that the \ac{SLSP} density closely follows the density of the total density and the end-particle density.
This is expected from the partition function, where the local \ac{SLSP} density depends linearly on the particle density\cite{chappa2012translationally}.
This is notable because \acp{SLSP} can only be created or deleted at the ends of the polymer chains.
However, there are small shifts near the onset of the interface.
The reflection of polymer conformations leads to an enrichment of polymer-end particles near the interface.
Similarly, we observe that the interface is slightly depleted of \acp{SLSP}.
There are two effects at work: i) The position of the \ac{SLSP} is defined as the midpoint between two bonded particles; a bond perpendicular to the interface cannot come closer to the interface than half a bond length.
This shifts the apparent onset of the \ac{SLSP} density away from the particle interface.
Further, ii), since particles close to the interface have less neighbors to potentially form a \ac{SLSP}, the partition function predicts a reduction of \ac{SLSP}s.
However, these effects are small $\ll \sigma$, which is only approximately represented in a highly coarse-grained model.
The particle density at the end of the polymer, combined with fewer \acp{SLSP} close to the interface, could potentially trap entanglements because it is likely that a \ac{SLSP} slides to the end of a polymer, where it can be destroyed and the entanglement is released.
We do not observe this dynamics in our droplet deposition simulations.
In the Supplementary Material, we present similar density profiles for the polymer droplet; the effect is less pronounced in that example.

\section{Conclusion}

We have proposed an extension of the \ac{SLSP} model from \citeauthor{chappa2012translationally} \cite{chappa2012translationally} that is capable of describing the dynamics of entangled polymers in the presence of explicit liquid-vapor or liquid-solid interfaces.
The \ac{MDPD} non-bonded interaction adopted here allows for liquid-vapor co-existence, with only minor additional computational demands.
The \ac{SLSP} model is compatible with this choice, as long as a compensating potential is used to correct for the attractive effects of additional springs in the bulk phase.
This combination enables study of high molecular weight polymers in a setting where chain conformations are constrained by a liquid-vapor interface, where entanglements play a key role.
The usefulness of the model has been illustrated by describing the deposition of polymer droplets on a substrate, where we have shown that the wetting process strongly depends on the degree of entanglement in the material.

More generally, the model proposed here could provide a foundation for the development of digital twins of experimentally relevant systems, including micro-rheometers based on droplet deformation.
With droplets, microscopic amounts of a sample are sufficient to probe the rheological properties of entangled melts using, for example, ultrafast imaging or atomic force microscopy \ac{AFM}.
With this new model, one could predict bulk rheology by relying on measurements on microscopic samples.
Furthermore, the use of an explicit liquid-vapor interface also allows for the modeling of evaporation, thereby enabling combined studies of entangled dynamics during non-equilibrium evaporation processes.

\begin{acknowledgement}

This work is supported by the Department of Energy, Basic Energy Sciences, and the Division of Materials Science and Engineering.
This work was completed in part with computational resources provided by the University of Chicago's \ac{RCC}.
The authors thank Kai-Uwe Hollborn, supervised by Marcus M\"uller, for his contribution to the \ac{SLSP} plugin.
For fruitful discussions, we thank
%Heyi Liang, Joshua Mysona, and
Juhae Park.
For experimental inspiration and discussion, we thank Carolin Wahl, Allen Guo, Jordan Swisher, and Chad Mirkin.

\end{acknowledgement}

\begin{suppinfo}

Additional information is available, provide further figures, demonstrating the importance of the compensating potential in case of liquid-vapor interfaces and further illustrations.

\end{suppinfo}

\bibliography{references}

\providecommand{\latin}[1]{#1}
\makeatletter
\providecommand{\doi}
  {\begingroup\let\do\@makeother\dospecials
  \catcode`\{=1 \catcode`\}=2 \doi@aux}
\providecommand{\doi@aux}[1]{\endgroup\texttt{#1}}
\makeatother
\providecommand*\mcitethebibliography{\thebibliography}
\csname @ifundefined\endcsname{endmcitethebibliography}
  {\let\endmcitethebibliography\endthebibliography}{}
\begin{mcitethebibliography}{45}
\providecommand*\natexlab[1]{#1}
\providecommand*\mciteSetBstSublistMode[1]{}
\providecommand*\mciteSetBstMaxWidthForm[2]{}
\providecommand*\mciteBstWouldAddEndPuncttrue
  {\def\EndOfBibitem{\unskip.}}
\providecommand*\mciteBstWouldAddEndPunctfalse
  {\let\EndOfBibitem\relax}
\providecommand*\mciteSetBstMidEndSepPunct[3]{}
\providecommand*\mciteSetBstSublistLabelBeginEnd[3]{}
\providecommand*\EndOfBibitem{}
\mciteSetBstSublistMode{f}
\mciteSetBstMaxWidthForm{subitem}{(\alph{mcitesubitemcount})}
\mciteSetBstSublistLabelBeginEnd
  {\mcitemaxwidthsubitemform\space}
  {\relax}
  {\relax}

\bibitem[De~Gennes and Leger(1982)De~Gennes, and Leger]{de1982dynamics}
De~Gennes,~P.; Leger,~L. Dynamics of entangled polymer chains. \emph{Annual
  Review of Physical Chemistry} \textbf{1982}, \emph{33}, 49--61\relax
\mciteBstWouldAddEndPuncttrue
\mciteSetBstMidEndSepPunct{\mcitedefaultmidpunct}
{\mcitedefaultendpunct}{\mcitedefaultseppunct}\relax
\EndOfBibitem
\bibitem[Gillies and Prestidge(2004)Gillies, and
  Prestidge]{gillies2004interaction}
Gillies,~G.; Prestidge,~C.~A. Interaction forces, deformation and nano-rheology
  of emulsion droplets as determined by colloid probe AFM. \emph{Advances in
  colloid and interface science} \textbf{2004}, \emph{108}, 197--205\relax
\mciteBstWouldAddEndPuncttrue
\mciteSetBstMidEndSepPunct{\mcitedefaultmidpunct}
{\mcitedefaultendpunct}{\mcitedefaultseppunct}\relax
\EndOfBibitem
\bibitem[Waigh(2016)]{waigh2016advances}
Waigh,~T.~A. Advances in the microrheology of complex fluids. \emph{Reports on
  Progress in Physics} \textbf{2016}, \emph{79}, 074601\relax
\mciteBstWouldAddEndPuncttrue
\mciteSetBstMidEndSepPunct{\mcitedefaultmidpunct}
{\mcitedefaultendpunct}{\mcitedefaultseppunct}\relax
\EndOfBibitem
\bibitem[Chappa \latin{et~al.}(2012)Chappa, Morse, Zippelius, and
  M\"uller]{chappa2012translationally}
Chappa,~V.; Morse,~D.~C.; Zippelius,~A.; M\"uller,~M. Translationally Invariant
  Slip-Spring Model for Entangled Polymer Dynamics. \emph{Phys. Rev. Lett.}
  \textbf{2012}, \emph{109}, 148302\relax
\mciteBstWouldAddEndPuncttrue
\mciteSetBstMidEndSepPunct{\mcitedefaultmidpunct}
{\mcitedefaultendpunct}{\mcitedefaultseppunct}\relax
\EndOfBibitem
\bibitem[Uneyama and Masubuchi(2012)Uneyama, and Masubuchi]{uneyama2012multi}
Uneyama,~T.; Masubuchi,~Y. Multi-chain slip-spring model for entangled polymer
  dynamics. \emph{The Journal of chemical physics} \textbf{2012}, \emph{137},
  154902\relax
\mciteBstWouldAddEndPuncttrue
\mciteSetBstMidEndSepPunct{\mcitedefaultmidpunct}
{\mcitedefaultendpunct}{\mcitedefaultseppunct}\relax
\EndOfBibitem
\bibitem[Ram{\ifmmode\acute{\imath}\else\'{\i}\fi}rez-Hern{\ifmmode\acute{a}\else\'{a}\fi}ndez
  \latin{et~al.}(2013)Ram{\ifmmode\acute{\imath}\else\'{\i}\fi}rez-Hern{\ifmmode\acute{a}\else\'{a}\fi}ndez,
  M{\ifmmode\ddot{u}\else\"{u}\fi}ller, and de~Pablo]{Hernandez2013Jan}
Ram{\ifmmode\acute{\imath}\else\'{\i}\fi}rez-Hern{\ifmmode\acute{a}\else\'{a}\fi}ndez,~A.;
  M{\ifmmode\ddot{u}\else\"{u}\fi}ller,~M.; de~Pablo,~J.~J. {Theoretically
  informed entangled polymer simulations: linear and non-linear rheology of
  melts}. \emph{Soft Matter} \textbf{2013}, \emph{9}, 2030--2036\relax
\mciteBstWouldAddEndPuncttrue
\mciteSetBstMidEndSepPunct{\mcitedefaultmidpunct}
{\mcitedefaultendpunct}{\mcitedefaultseppunct}\relax
\EndOfBibitem
\bibitem[Ram{\ifmmode\acute{\imath}\else\'{\i}\fi}rez-Hern{\ifmmode\acute{a}\else\'{a}\fi}ndez
  \latin{et~al.}(2013)Ram{\ifmmode\acute{\imath}\else\'{\i}\fi}rez-Hern{\ifmmode\acute{a}\else\'{a}\fi}ndez,
  Detcheverry, Peters, Chappa, Schweizer, M{\ifmmode\ddot{u}\else\"{u}\fi}ller,
  and de~Pablo]{Hernandez2013Aug}
Ram{\ifmmode\acute{\imath}\else\'{\i}\fi}rez-Hern{\ifmmode\acute{a}\else\'{a}\fi}ndez,~A.;
  Detcheverry,~F.~A.; Peters,~B.~L.; Chappa,~V.~C.; Schweizer,~K.~S.;
  M{\ifmmode\ddot{u}\else\"{u}\fi}ller,~M.; de~Pablo,~J.~J. {Dynamical
  Simulations of Coarse Grain Polymeric Systems: Rouse and Entangled Dynamics}.
  \emph{Macromolecules} \textbf{2013}, \emph{46}, 6287--6299\relax
\mciteBstWouldAddEndPuncttrue
\mciteSetBstMidEndSepPunct{\mcitedefaultmidpunct}
{\mcitedefaultendpunct}{\mcitedefaultseppunct}\relax
\EndOfBibitem
\bibitem[Ramirez-Hernandez \latin{et~al.}(2017)Ramirez-Hernandez, Peters,
  Schneider, Andreev, Schieber, M\"uller, and de~Pablo]{ramirez2017multi}
Ramirez-Hernandez,~A.; Peters,~B.~L.; Schneider,~L.; Andreev,~M.;
  Schieber,~J.~D.; M\"uller,~M.; de~Pablo,~J.~J. A multi-chain polymer
  slip-spring model with fluctuating number of entanglements: Density
  fluctuations, confinement, and phase separation. \emph{The Journal of
  Chemical Physics} \textbf{2017}, \emph{146}, 014903\relax
\mciteBstWouldAddEndPuncttrue
\mciteSetBstMidEndSepPunct{\mcitedefaultmidpunct}
{\mcitedefaultendpunct}{\mcitedefaultseppunct}\relax
\EndOfBibitem
\bibitem[Sgouros \latin{et~al.}(2017)Sgouros, Megariotis, and
  Theodorou]{sgouros2017slip}
Sgouros,~A.; Megariotis,~G.; Theodorou,~D. Slip-spring model for the linear and
  nonlinear viscoelastic properties of molten polyethylene derived from
  atomistic simulations. \emph{Macromolecules} \textbf{2017}, \emph{50},
  4524--4541\relax
\mciteBstWouldAddEndPuncttrue
\mciteSetBstMidEndSepPunct{\mcitedefaultmidpunct}
{\mcitedefaultendpunct}{\mcitedefaultseppunct}\relax
\EndOfBibitem
\bibitem[Vogiatzis \latin{et~al.}(2017)Vogiatzis, Megariotis, and
  Theodorou]{vogiatzis2017equation}
Vogiatzis,~G.~G.; Megariotis,~G.; Theodorou,~D.~N. Equation of state based slip
  spring model for entangled polymer dynamics. \emph{Macromolecules}
  \textbf{2017}, \emph{50}, 3004--3029\relax
\mciteBstWouldAddEndPuncttrue
\mciteSetBstMidEndSepPunct{\mcitedefaultmidpunct}
{\mcitedefaultendpunct}{\mcitedefaultseppunct}\relax
\EndOfBibitem
\bibitem[Masubuchi(2018)]{masubuchi2018multichain}
Masubuchi,~Y. Multichain slip-spring simulations for branch polymers.
  \emph{Macromolecules} \textbf{2018}, \emph{51}, 10184--10193\relax
\mciteBstWouldAddEndPuncttrue
\mciteSetBstMidEndSepPunct{\mcitedefaultmidpunct}
{\mcitedefaultendpunct}{\mcitedefaultseppunct}\relax
\EndOfBibitem
\bibitem[Langeloth \latin{et~al.}(2013)Langeloth, Masubuchi, B{\"o}hm, and
  M{\"u}ller-Plathe]{langeloth2013recovering}
Langeloth,~M.; Masubuchi,~Y.; B{\"o}hm,~M.~C.; M{\"u}ller-Plathe,~F. Recovering
  the reptation dynamics of polymer melts in dissipative particle dynamics
  simulations via slip-springs. \emph{The Journal of chemical physics}
  \textbf{2013}, \emph{138}, 104907\relax
\mciteBstWouldAddEndPuncttrue
\mciteSetBstMidEndSepPunct{\mcitedefaultmidpunct}
{\mcitedefaultendpunct}{\mcitedefaultseppunct}\relax
\EndOfBibitem
\bibitem[Masubuchi \latin{et~al.}(2016)Masubuchi, Langeloth, Böhm, and
  Inoue]{masubuchi2016multichain}
Masubuchi,~Y.; Langeloth,~M.; Böhm,~M.~C.; Inoue,~F.,~Tadashi Müller-Plathe A
  multichain slip-spring dissipative particle dynamics simulation method for
  entangled polymer solutions. \emph{Macromolecules} \textbf{2016}, \emph{49},
  9186--9191\relax
\mciteBstWouldAddEndPuncttrue
\mciteSetBstMidEndSepPunct{\mcitedefaultmidpunct}
{\mcitedefaultendpunct}{\mcitedefaultseppunct}\relax
\EndOfBibitem
\bibitem[Megariotis \latin{et~al.}(2018)Megariotis, Vogiatzis, Sgouros, and
  Theodorou]{megariotis2018slip}
Megariotis,~G.; Vogiatzis,~G.~G.; Sgouros,~A.~P.; Theodorou,~D.~N. Slip
  spring-based mesoscopic simulations of polymer networks: Methodology and the
  corresponding computational code. \emph{Polymers} \textbf{2018}, \emph{10},
  1156\relax
\mciteBstWouldAddEndPuncttrue
\mciteSetBstMidEndSepPunct{\mcitedefaultmidpunct}
{\mcitedefaultendpunct}{\mcitedefaultseppunct}\relax
\EndOfBibitem
\bibitem[Behbahani \latin{et~al.}(2021)Behbahani, Schneider, Rissanou,
  Chazirakis, Bacova, Jana, Li, Doxastakis, Polinska, Burkhart, \latin{et~al.}
  others]{behbahani2021dynamics}
Behbahani,~A.~F.; Schneider,~L.; Rissanou,~A.; Chazirakis,~A.; Bacova,~P.;
  Jana,~P.~K.; Li,~W.; Doxastakis,~M.; Polinska,~P.; Burkhart,~C.,
  \latin{et~al.}  Dynamics and rheology of polymer melts via hierarchical
  atomistic, coarse-grained, and slip-spring simulations. \emph{Macromolecules}
  \textbf{2021}, \emph{54}, 2740--2762\relax
\mciteBstWouldAddEndPuncttrue
\mciteSetBstMidEndSepPunct{\mcitedefaultmidpunct}
{\mcitedefaultendpunct}{\mcitedefaultseppunct}\relax
\EndOfBibitem
\bibitem[Schneider \latin{et~al.}(2021)Schneider, Fleck, Karimi-Varzaneh, and
  M\"uller-Plathe]{schneider2021simulation}
Schneider,~J.; Fleck,~F.; Karimi-Varzaneh,~H.~A.; M\"uller-Plathe,~F.
  Simulation of Elastomers by Slip-Spring Dissipative Particle Dynamics.
  \emph{Macromolecules} \textbf{2021}, \emph{54}, 5155--5166\relax
\mciteBstWouldAddEndPuncttrue
\mciteSetBstMidEndSepPunct{\mcitedefaultmidpunct}
{\mcitedefaultendpunct}{\mcitedefaultseppunct}\relax
\EndOfBibitem
\bibitem[Li \latin{et~al.}(2021)Li, Jana, Behbahani, Kritikos, Schneider,
  Polinska, Burkhart, Harmandaris, M\"uller, and Doxastakis]{li2021dynamics}
Li,~W.; Jana,~P.~K.; Behbahani,~A.~F.; Kritikos,~G.; Schneider,~L.;
  Polinska,~P.; Burkhart,~C.; Harmandaris,~V.~A.; M\"uller,~M.; Doxastakis,~M.
  Dynamics of Long Entangled Polyisoprene Melts via Multiscale Modeling.
  \emph{Macromolecules} \textbf{2021}, \emph{54}, 8693--8713\relax
\mciteBstWouldAddEndPuncttrue
\mciteSetBstMidEndSepPunct{\mcitedefaultmidpunct}
{\mcitedefaultendpunct}{\mcitedefaultseppunct}\relax
\EndOfBibitem
\bibitem[Megariotis \latin{et~al.}(2016)Megariotis, Vogiatzis, Schneider,
  M{\"u}ller, and Theodorou]{megariotis2016mesoscopic}
Megariotis,~G.; Vogiatzis,~G.~G.; Schneider,~L.; M{\"u}ller,~M.;
  Theodorou,~D.~N. Mesoscopic simulations of crosslinked polymer networks.
  \emph{J. Phys. Conf.} \textbf{2016}, \emph{738}, 012063\relax
\mciteBstWouldAddEndPuncttrue
\mciteSetBstMidEndSepPunct{\mcitedefaultmidpunct}
{\mcitedefaultendpunct}{\mcitedefaultseppunct}\relax
\EndOfBibitem
\bibitem[Hollborn \latin{et~al.}(2022)Hollborn, Schneider, and
  M{\"u}ller]{hollborn2022effect}
Hollborn,~K.-U.; Schneider,~L.; M{\"u}ller,~M. Effect of Slip-Spring Parameters
  on the Dynamics and Rheology of Soft, Coarse-Grained Polymer Models.
  \emph{The Journal of Physical Chemistry B} \textbf{2022}, \relax
\mciteBstWouldAddEndPunctfalse
\mciteSetBstMidEndSepPunct{\mcitedefaultmidpunct}
{}{\mcitedefaultseppunct}\relax
\EndOfBibitem
\bibitem[M\"uller(2011)]{muller2011studying}
M\"uller,~M. Studying Amphiphilic Self-assembly with Soft Coarse-Grained
  Models. \emph{J. Stat. Phys.} \textbf{2011}, \emph{145}, 967--1016\relax
\mciteBstWouldAddEndPuncttrue
\mciteSetBstMidEndSepPunct{\mcitedefaultmidpunct}
{\mcitedefaultendpunct}{\mcitedefaultseppunct}\relax
\EndOfBibitem
\bibitem[Tsch{\"o}p \latin{et~al.}(1998)Tsch{\"o}p, Kremer, Batoulis,
  B{\"u}rger, and Hahn]{tschop1998simulation}
Tsch{\"o}p,~W.; Kremer,~K.; Batoulis,~J.; B{\"u}rger,~T.; Hahn,~O. Simulation
  of polymer melts. I. Coarse-graining procedure for polycarbonates. \emph{Acta
  Polym.} \textbf{1998}, \emph{49}, 61--74\relax
\mciteBstWouldAddEndPuncttrue
\mciteSetBstMidEndSepPunct{\mcitedefaultmidpunct}
{\mcitedefaultendpunct}{\mcitedefaultseppunct}\relax
\EndOfBibitem
\bibitem[Harmandaris \latin{et~al.}(2006)Harmandaris, Adhikari, van~der Vegt,
  and Kremer]{harmandaris2006hierarchical}
Harmandaris,~V.; Adhikari,~N.; van~der Vegt,~N.~F.; Kremer,~K. Hierarchical
  modeling of polystyrene: From atomistic to coarse-grained simulations.
  \emph{Macromolecules} \textbf{2006}, \emph{39}, 6708--6719\relax
\mciteBstWouldAddEndPuncttrue
\mciteSetBstMidEndSepPunct{\mcitedefaultmidpunct}
{\mcitedefaultendpunct}{\mcitedefaultseppunct}\relax
\EndOfBibitem
\bibitem[Praprotnik \latin{et~al.}(2008)Praprotnik, delle Site, and
  Kremer]{Praprotnik08}
Praprotnik,~M.; delle Site,~L.; Kremer,~K. Multiscale Simulation of Soft
  Matter: from Scale Bridging To Adaptive Resolution. \emph{Ann. Rev. Phys.
  Chem.} \textbf{2008}, \emph{59}, 545--571\relax
\mciteBstWouldAddEndPuncttrue
\mciteSetBstMidEndSepPunct{\mcitedefaultmidpunct}
{\mcitedefaultendpunct}{\mcitedefaultseppunct}\relax
\EndOfBibitem
\bibitem[Fritz \latin{et~al.}(2009)Fritz, Harmandaris, Kremer, and van~der
  Vegt]{fritz2009coarse}
Fritz,~D.; Harmandaris,~V.~A.; Kremer,~K.; van~der Vegt,~N.~F. Coarse-grained
  polymer melts based on isolated atomistic chains: Simulation of polystyrene
  of different tacticities. \emph{Macromolecules} \textbf{2009}, \emph{42},
  7579--7588\relax
\mciteBstWouldAddEndPuncttrue
\mciteSetBstMidEndSepPunct{\mcitedefaultmidpunct}
{\mcitedefaultendpunct}{\mcitedefaultseppunct}\relax
\EndOfBibitem
\bibitem[Padding and Briels(2011)Padding, and Briels]{padding2011systematic}
Padding,~J.; Briels,~W.~J. Systematic coarse-graining of the dynamics of
  entangled polymer melts: the road from chemistry to rheology. \emph{J. Phys.
  Condens. Matter} \textbf{2011}, \emph{23}, 233101\relax
\mciteBstWouldAddEndPuncttrue
\mciteSetBstMidEndSepPunct{\mcitedefaultmidpunct}
{\mcitedefaultendpunct}{\mcitedefaultseppunct}\relax
\EndOfBibitem
\bibitem[Webb \latin{et~al.}(2018)Webb, Delannoy, and De~Pablo]{webb2018graph}
Webb,~M.~A.; Delannoy,~J.-Y.; De~Pablo,~J.~J. Graph-based approach to
  systematic molecular coarse-graining. \emph{J. Chem. Theory Comput.}
  \textbf{2018}, \emph{15}, 1199--1208\relax
\mciteBstWouldAddEndPuncttrue
\mciteSetBstMidEndSepPunct{\mcitedefaultmidpunct}
{\mcitedefaultendpunct}{\mcitedefaultseppunct}\relax
\EndOfBibitem
\bibitem[Dhamankar and Webb(2021)Dhamankar, and Webb]{dhamankar2021chemically}
Dhamankar,~S.; Webb,~M.~A. Chemically specific coarse-graining of polymers:
  Methods and prospects. \emph{J. Polym. Sci.} \textbf{2021}, \emph{59},
  2613--2643\relax
\mciteBstWouldAddEndPuncttrue
\mciteSetBstMidEndSepPunct{\mcitedefaultmidpunct}
{\mcitedefaultendpunct}{\mcitedefaultseppunct}\relax
\EndOfBibitem
\bibitem[Doi and Edwards(1988)Doi, and Edwards]{doi1988the}
Doi,~M.; Edwards,~S. \emph{The Theory of Polymer Dynamics}; 1988\relax
\mciteBstWouldAddEndPuncttrue
\mciteSetBstMidEndSepPunct{\mcitedefaultmidpunct}
{\mcitedefaultendpunct}{\mcitedefaultseppunct}\relax
\EndOfBibitem
\bibitem[Ramirez-Hernandez \latin{et~al.}(2018)Ramirez-Hernandez, Peters,
  Schneider, Andreev, Schieber, M\"uller, Kr\"oger, and
  de~Pablo]{ramirez2018detailed}
Ramirez-Hernandez,~A.; Peters,~B.~L.; Schneider,~L.; Andreev,~M.;
  Schieber,~J.~D.; M\"uller,~M.; Kr\"oger,~M.; de~Pablo,~J.~J. A detailed
  examination of the topological constraints of lamellae-forming block
  copolymers. \emph{Macromolecules} \textbf{2018}, \emph{51}, 2110--2124\relax
\mciteBstWouldAddEndPuncttrue
\mciteSetBstMidEndSepPunct{\mcitedefaultmidpunct}
{\mcitedefaultendpunct}{\mcitedefaultseppunct}\relax
\EndOfBibitem
\bibitem[Liang \latin{et~al.}(2022)Liang, Yoshimoto, Gil, Kitabata, Yamamoto,
  and de~Pablo]{liang2022bottom}
Liang,~H.; Yoshimoto,~K.; Gil,~P.; Kitabata,~M.; Yamamoto,~U.; de~Pablo,~J.~J.
  Bottom-Up Multiscale Approach to Estimate Viscoelastic Properties of
  Entangled Polymer Melts with High Glass Transition Temperature.
  \emph{Macromolecules} \textbf{2022}, \emph{55}, 3159--3165\relax
\mciteBstWouldAddEndPuncttrue
\mciteSetBstMidEndSepPunct{\mcitedefaultmidpunct}
{\mcitedefaultendpunct}{\mcitedefaultseppunct}\relax
\EndOfBibitem
\bibitem[Dreyer \latin{et~al.}(2022)Dreyer, Ibbeken, Schneider, Blagojevic,
  Radjabian, Abetz, and M{\"u}ller]{dreyer2022simulation}
Dreyer,~O.; Ibbeken,~G.; Schneider,~L.; Blagojevic,~N.; Radjabian,~M.;
  Abetz,~V.; M{\"u}ller,~M. Simulation of Solvent Evaporation from a Diblock
  Copolymer Film: Orientation of the Cylindrical Mesophase.
  \emph{Macromolecules} \textbf{2022}, \relax
\mciteBstWouldAddEndPunctfalse
\mciteSetBstMidEndSepPunct{\mcitedefaultmidpunct}
{}{\mcitedefaultseppunct}\relax
\EndOfBibitem
\bibitem[Michman \latin{et~al.}(2019)Michman, Langenberg, Stenger, Oded,
  Schvartzman, M{\"u}ller, and Shenhar]{michman2019controlled}
Michman,~E.; Langenberg,~M.; Stenger,~R.; Oded,~M.; Schvartzman,~M.;
  M{\"u}ller,~M.; Shenhar,~R. Controlled spacing between nanopatterned regions
  in block copolymer films obtained by utilizing substrate topography for local
  film thickness differentiation. \emph{ACS applied materials \& interfaces}
  \textbf{2019}, \emph{11}, 35247--35254\relax
\mciteBstWouldAddEndPuncttrue
\mciteSetBstMidEndSepPunct{\mcitedefaultmidpunct}
{\mcitedefaultendpunct}{\mcitedefaultseppunct}\relax
\EndOfBibitem
\bibitem[Groot and Warren(1997)Groot, and Warren]{groot1997dissipative}
Groot,~R.~D.; Warren,~P.~B. Dissipative particle dynamics: Bridging the gap
  between atomistic and mesoscopic simulation. \emph{J. Chem. Phys.}
  \textbf{1997}, \emph{107}, 4423--4435\relax
\mciteBstWouldAddEndPuncttrue
\mciteSetBstMidEndSepPunct{\mcitedefaultmidpunct}
{\mcitedefaultendpunct}{\mcitedefaultseppunct}\relax
\EndOfBibitem
\bibitem[Warren and Espanol(1995)Warren, and Espanol]{espanol1995statistical}
Warren,~P.; Espanol,~P. Statistical-mechanics of dissipative particle dynamics.
  \emph{EPL (Europhys. Lett.)} \textbf{1995}, \emph{30}, 191196\relax
\mciteBstWouldAddEndPuncttrue
\mciteSetBstMidEndSepPunct{\mcitedefaultmidpunct}
{\mcitedefaultendpunct}{\mcitedefaultseppunct}\relax
\EndOfBibitem
\bibitem[Warren(2003)]{warren2003vapor}
Warren,~P. Vapor-liquid coexistence in many-body dissipative particle dynamics.
  \emph{Physical Review E} \textbf{2003}, \emph{68}, 066702\relax
\mciteBstWouldAddEndPuncttrue
\mciteSetBstMidEndSepPunct{\mcitedefaultmidpunct}
{\mcitedefaultendpunct}{\mcitedefaultseppunct}\relax
\EndOfBibitem
\bibitem[Warren(2013)]{warren2013no}
Warren,~P.~B. No-go theorem in many-body dissipative particle dynamics.
  \emph{Physical Review E} \textbf{2013}, \emph{87}, 045303\relax
\mciteBstWouldAddEndPuncttrue
\mciteSetBstMidEndSepPunct{\mcitedefaultmidpunct}
{\mcitedefaultendpunct}{\mcitedefaultseppunct}\relax
\EndOfBibitem
\bibitem[Zhao \latin{et~al.}(2021)Zhao, Chen, Zhang, and Liu]{zhao2021review}
Zhao,~J.; Chen,~S.; Zhang,~K.; Liu,~Y. A review of many-body dissipative
  particle dynamics (MDPD): Theoretical models and its applications.
  \emph{Physics of Fluids} \textbf{2021}, \emph{33}, 112002\relax
\mciteBstWouldAddEndPuncttrue
\mciteSetBstMidEndSepPunct{\mcitedefaultmidpunct}
{\mcitedefaultendpunct}{\mcitedefaultseppunct}\relax
\EndOfBibitem
\bibitem[Anderson \latin{et~al.}(2008)Anderson, Lorenz, and
  Travesset]{anderson2008general}
Anderson,~J.~A.; Lorenz,~C.~D.; Travesset,~A. General Purpose Molecular
  Dynamics Simulations Fully Implemented on Graphics Processing Units.
  \emph{Journal of Computational Physics} \textbf{2008}, \emph{227}, 5342 --
  5359\relax
\mciteBstWouldAddEndPuncttrue
\mciteSetBstMidEndSepPunct{\mcitedefaultmidpunct}
{\mcitedefaultendpunct}{\mcitedefaultseppunct}\relax
\EndOfBibitem
\bibitem[Phillips \latin{et~al.}(2011)Phillips, Anderson, and
  Glotzer]{phillips2011pseudo}
Phillips,~C.~L.; Anderson,~J.~A.; Glotzer,~S.~C. Pseudo-Random Number
  Generation for Brownian Dynamics and Dissipative Particle Dynamics
  Simulations on GPU Devices. \emph{Journal of Computational Physics}
  \textbf{2011}, \emph{230}, 7191--7201\relax
\mciteBstWouldAddEndPuncttrue
\mciteSetBstMidEndSepPunct{\mcitedefaultmidpunct}
{\mcitedefaultendpunct}{\mcitedefaultseppunct}\relax
\EndOfBibitem
\bibitem[Glaser \latin{et~al.}(2015)Glaser, Nguyen, Anderson, Lui, Spiga,
  Millan, Morse, and Glotzer]{glaser2015strong}
Glaser,~J.; Nguyen,~T.~D.; Anderson,~J.~A.; Lui,~P.; Spiga,~F.; Millan,~J.~A.;
  Morse,~D.~C.; Glotzer,~S.~C. Strong scaling of general-purpose molecular
  dynamics simulations on GPUs. \emph{Computer Physics Communications}
  \textbf{2015}, \emph{192}, 97--107\relax
\mciteBstWouldAddEndPuncttrue
\mciteSetBstMidEndSepPunct{\mcitedefaultmidpunct}
{\mcitedefaultendpunct}{\mcitedefaultseppunct}\relax
\EndOfBibitem
\bibitem[Anderson \latin{et~al.}(2020)Anderson, Glaser, and
  Glotzer]{anderson2020hoomd}
Anderson,~J.~A.; Glaser,~J.; Glotzer,~S.~C. HOOMD-blue: A Python package for
  high-performance molecular dynamics and hard particle Monte Carlo
  simulations. \emph{Computational Materials Science} \textbf{2020},
  \emph{173}, 109363\relax
\mciteBstWouldAddEndPuncttrue
\mciteSetBstMidEndSepPunct{\mcitedefaultmidpunct}
{\mcitedefaultendpunct}{\mcitedefaultseppunct}\relax
\EndOfBibitem
\bibitem[Jana \latin{et~al.}(2022)Jana, Bačová, Schneider, Kobayashi,
  Hollborn, Polińska, Burkhart, Harmandaris, and Müller]{kumar2022wall}
Jana,~P.~K.; Bačová,~P.; Schneider,~L.; Kobayashi,~H.; Hollborn,~K.-U.;
  Polińska,~P.; Burkhart,~C.; Harmandaris,~V.~A.; Müller,~M. Wall-Spring
  Thermostat: A Novel Approach for Controlling the Dynamics of Soft
  Coarse-Grained Polymer Fluids at Surfaces. \emph{Macromolecules}
  \textbf{2022}, \emph{55}, 5550--5566\relax
\mciteBstWouldAddEndPuncttrue
\mciteSetBstMidEndSepPunct{\mcitedefaultmidpunct}
{\mcitedefaultendpunct}{\mcitedefaultseppunct}\relax
\EndOfBibitem
\bibitem[Liu \latin{et~al.}(2020)Liu, Petrosko, Zheng, and
  Mirkin]{liu2020evolution}
Liu,~G.; Petrosko,~S.~H.; Zheng,~Z.; Mirkin,~C.~A. Evolution of dip-pen
  nanolithography (DPN): From molecular patterning to materials discovery.
  \emph{Chemical reviews} \textbf{2020}, \emph{120}, 6009--6047\relax
\mciteBstWouldAddEndPuncttrue
\mciteSetBstMidEndSepPunct{\mcitedefaultmidpunct}
{\mcitedefaultendpunct}{\mcitedefaultseppunct}\relax
\EndOfBibitem
\bibitem[Silberberg(1982)]{silberberg1982distribution}
Silberberg,~A. Distribution of conformations and chain ends near the surface of
  a melt of linear flexible macromolecules. \emph{Journal of Colloid and
  Interface Science} \textbf{1982}, \emph{90}, 86--91\relax
\mciteBstWouldAddEndPuncttrue
\mciteSetBstMidEndSepPunct{\mcitedefaultmidpunct}
{\mcitedefaultendpunct}{\mcitedefaultseppunct}\relax
\EndOfBibitem
\end{mcitethebibliography}

\end{document}

\section{Importance of compensating potential}

The compensating potential of the \ac{SLSP}-model\cite{chappa2012translationally} ensures that the liquid-vapor and liquid-wall interface is independent of the density of \ac{SLSP} (\autoref{fig:profile}).
This ensures that dynamic and static properties are separated and can be tuned independently during the top-down coarse-graining process. The system presented is identical to the system shown in Figure 4 of the main manuscript.

\begin{figure}
    \centering
    \begin{subfigure}{0.5\multifigurewidth}
    \includegraphics[width=\textwidth]{figures/profile.pdf}
    \caption{Liquid-vapor interface for differing number of entanglements: without the compensating potential interface position and shape changes.}
    \label{fig:profile-vapor}
    \end{subfigure}
    \begin{subfigure}{0.5\multifigurewidth}
    \includegraphics[width=\textwidth]{figures/wall_profile.pdf}
        \caption{Wall-liquid interface for differing number of entanglements: without the compensating potential the packing close to the wall slightly changes.}
        \label{fig:profile-wall}
    \end{subfigure}
    \caption{Density profile of a liquid-vapor system with (top) and without (bottom) the compensating potential.
             In different colors, we represent the density of differently entangled materials, by changing the fugacity $z$.
             It becomes apparent that the compensating potential is necessary for this model to avoid undesired effects near interfaces.}
    \label{fig:profile}
\end{figure}
\section{\ac{SLSP} interface profile}

We present with \autoref{fig:slsp-prof} profiles of the density, polymer-end-particle density and \acp{SLSP} for the polymer droplet as deposited on the surface.

\begin{figure}
    \centering
    \begin{subfigure}{0.5\multifigurewidth}
    \includegraphics[width=\textwidth]{figures/profile_drop.pdf}
    \caption{Spherical non-wetting droplet with $\epsilon_\text{LJ} = k_B T$.}
    \label{fig:sphere-prof}
    \end{subfigure}
    \begin{subfigure}{0.5\multifigurewidth}
    \includegraphics[width=\textwidth]{figures/profile_slsp.pdf}
    \caption{Wetting droplet at $\epsilon_\text{LJ} = 8.5 k_BT$ at $t=3000\tau$}
    \label{fig:wet-prof}
    \end{subfigure}
    \caption{We present the normalized probability distributions of particle, polymer-end-particles and \acp{SLSP}. The main panel shows the distribution as function of the radius of the droplet appropriate for the initial spherical form. The inset shows the distribution as function of radius $R_{yz}$ in the plane parallel to the substrate appropriate for the later state.}
    \label{fig:slsp-prof}
\end{figure}

\clearpage
\bibliography{references}